\documentclass[prl,twocolumn,showpacs,preprintnumbers,amsmath,amssymb]{revtex4}
\usepackage{graphicx}
\usepackage{color}
\usepackage{amsmath}
\usepackage{epstopdf}
\usepackage{amsbsy}
\usepackage{dcolumn}
\usepackage{bm}

\newcommand{\vS}{{\mbox{\boldmath$S$}}}

\newcommand{\vE}{\mbox{\boldmath$E$}}

\newcommand{\vsig}{\mbox{\boldmath$\sigma$}}

\newcommand{\mhx}{\hat{\mbox{\boldmath$x$}}}
\newcommand{\mhy}{\hat{\mbox{\boldmath$y$}}}
\newcommand{\mhz}{\hat{\mbox{\boldmath$z$}}}

\newcommand{\vH}{\mbox{\boldmath$H$}}

\newcommand{\vP}{\mbox{\boldmath$P$}}
\newcommand{\vM}{\mbox{\boldmath$M$}}

\newcommand{\vL}{\mbox{\boldmath$L$}}

\begin{document}

\title{Theory of the magnetoeletric effect in a lightly doped high-T$_c$ cuprate}
\author{S. Mukherjee$^{1,2,3}$, B. M. Andersen$^2$, Z. Viskadourakis$^1$, I. Radulov$^1$, C. Panagopoulos$^{1,4,5}$}
\affiliation{
$^1$ Institute of Electronic Structure and Laser, Foundation for Research and Technology Hellas, Heraklion, 70013, Greece\\
$^2$Niels Bohr Institute, University of Copenhagen, DK-2100 Copenhagen \O, Denmark\\
$^3$Niels Bohr International Academy, Niels Bohr Institute, University of Copenhagen, Blegdamsvej 17, DK-2100 Copenhagen \O, Denmark\\
$^4$Department of Physics, University of Crete, Heraklion, 71003, Greece\\
$^5$ Division of Physics and Applied Physics, Nanyang Technological University, 21 Nanyang Link, 637371, Singapore}
\date{\today{}}

\begin{abstract}
In a recent study Viskadourakis {\it et al.} discovered that extremely underdoped La$_2$CuO$_{4+x}$ is a relaxor ferroelectric and a magnetoelectric material at low temperatures. It is further observed that the magnetoelectric response is anisotropic for different directions of electric polarization and applied magnetic field. By constructing an appropriate Landau theory, we show that a bi-quadratic magnetoelectric coupling can explain the experimentally observed polarization dependence on magnetic field. This coupling leads to several novel low-temperature effects including a feedback enhancement of the magnetization below the ferroelectric transition, and a predicted magnetocapacitive effect.

\end{abstract}

\pacs{64.70.P-, 74.72.Cj, 77.80.-e, 77.80.Jk}
\maketitle

The field of magnetoelectrics has witnessed intense theoretical and experimental progress in the recent years, mainly driven by the discovery of new  materials exhibiting a non-linear magnetoelectric effect i.e. the dominant magnetoelectric coupling is of higher order than a bilinear coupling between electric and magnetic fields \cite{Eerenstein:2006,Cheong:2007}. Among the materials discovered with such unusual physical properties are the so-called \textit{birelaxors} \cite{Levstik:2007}. These systems show both relaxor ferroelectric and relaxor magnetic properties and are associated with spin-charge coupling at a mesoscopic scale. Focussing on the parent high-T$_c$ superconductor La$_2$CuO$_{4+x}$ (LCO) with exceptionally low carrier concentration $n=10^{17}\rm{cm}^{-3}$, we have recently found that this material is in fact a ferroelectric at low temperatures \cite{Zach:2011}. More specifically, LCO has been shown to be a relaxor ferroelectric where the dielectric mode behavior is caused by freezing of randomly oriented polarized regions \cite{Cross:1987,Samara:2003}. In addition, LCO exhibits a distinct magnetoelectric effect with a pronounced dependence of the polarization on an externally applied magnetic field. These results shine new light on the nature of impurities and doped charge carriers in antiferromagnetic Mott insulators. Here, we focus on the magnetoelectric effect discovered in Ref.~\onlinecite{Zach:2011}, and show how  non-linear terms coupling polarization and magnetization naturally lead to the observed field effect.

Parent cuprate superconductors are 2D antiferromagnets with weak interplanar exchange coupling giving rise to 3D long-range N\'{e}el order \cite{Kastner:1998,Keimer1:1992}. In the case of LCO, the Cu spins are slightly canted out of the CuO$_2$ planes because of a finite Dzyaloshinskii-Moriya (DM) interaction existing in the low temperature orthorhombic phase (LTO) \cite{Thio:1988}. Though this allows a small ferromagnetic moment to build up on each CuO$_2$ plane, the net magnetic moment is zero since the moments in consecutive layers are oriented in opposite directions. On application of an external magnetic field a first order spin flop transition is observed at a critical magnetic field $H_{\rm{sf}}\sim5\rm{T}$ \cite{Thio:1988,Reehuis:2006}. Clear evidence for coupled spin and charge degrees of freedom in these systems come from the observation of pronounced discontinuities in resistivity and dielectric constant at a magnetic field corresponding to $H_{\rm{sf}}$ \cite{Zach:2011, Thio:1990}. Further evidence of such coupling is found in LCO by the possibility to detwin these crystals e.g. by the application of an in-plane magnetic field \cite{Lavrov:2002}.

In the following we use a symmetry based analysis to identify the particular magnetoelectric interaction terms that are responsible for spin-charge coupling in underdoped LCO. We further construct a Landau theory, and show how this model reproduces all the qualitative features of the magnetic field dependence of the polarization curves reported in Ref.~\onlinecite{Zach:2011}.

Below approximately 530K the crystal structure of LCO  is LTO with space group Cmca (D$_{2h}^{18}$).
If $(\mhx,\mhy,\mhz)$ denote unit vectors along the crystal axes then we define fractional translations by
\begin{eqnarray}
\tau = {1\over 2}(a\mhx+c\mhz)
\tau'= {1\over 2}(a\mhx+b\mhy),
\end{eqnarray}
where $a,b,c$ are the lattice constants. The symmetry elements of this crystal structure are then written as $G=G_0+\tau G_0$, where $G_0$ contains the eight elements
\begin{eqnarray}
E,I,\sigma_a,\sigma_b',\sigma_c', C_{2a}, C_{2b}', C_{2c}'.
\end{eqnarray}
Here E denotes the identity, I inversion about a Cu site, $\sigma_a,\sigma_b,\sigma_c$ reflections about the planes $x=a/2, y=b/2, z=c/2$, and $C_{2a},C_{2b},C_{2c}$ are $180^{o}$ rotations about the axes that emanate from the center of the unit cell. Primed elements must be complemented by translation $\tau'$ that is itself not a symmetry operation.

Taking into account the above mentioned symmetry properties of the Cmca space group, the free energy can be expressed as a sum of three contributions
\begin{align}
F=F_M+F_{MP}+F_P.
\end{align}
Here, $F_M$ is the purely magnetic free energy, $F_{MP}$ is the magnetoelectric contribution, and $F_{P}$ is the polarization free energy.
The magnetic free energy that accounts for the crystal structure of the LTO phase has been studied previously by e.g. Thio {\it et al.} \cite{Thio:1994} and is given by
\begin{align}
& F_M = {1\over 2}\sum_{i=1}^2[{\chi_{2D}^{-1}\over 2}L_i^2+{1\over 4}AL_i^4+{1\over 6}BL_i^6-CL_iM_i\nonumber\\
 & +{\chi_0^{-1}\over 2}M_i^2 -H_cM_i-H_{ab}L_i]+{1\over 2}J_{\bot}L_1L_2.
\end{align}
Here, the out-of-plane ($c$ direction) [in-plane ($a-b$ plane)] applied magnetic field is represented by $H_c$ [$H_{ab}$]. The order parameter $M_i=(S_{Ai}+S_{Bi})/2$ is the ferromagnetic moment per spin with $S_{Ai},S_{Bi}$ being the sublattice spins in the $i^{th}$ plane, and $L_i=(S_{Ai}-S_{Bi})/2$ is the antiferromagnetic order parameter ($L_i||a$). The spins are slightly canted due to the DM interaction term $-CM_iL_i$, which causes them to lie in the $a-c$ plane of the magnetic unit cell. The coupling between the different planes is included by the $J_{\bot}$ term.

The presence of an inversion symmetry in the space group of the crystal forbids any linear magnetoelectric effect \cite{Dzyaloshinskii:1991} and the physics is dominated by non-linear coupling terms. We can focus on the largest non-linear terms by further noting that the experimentally observed polarization response is symmetric under inversion of the external magnetic field (i.e $\vP(\vH)=\vP(-\vH)$). This implies that the dominant couplings are of even order in the magnetic order parameter. Hence, the following terms contribute to the magnetoelectric coupling
\begin{align}
& F_{MP}= \sum_{\alpha,i}({\gamma_{1\alpha}\over 2}L_i^2+{\gamma_{2\alpha}\over 2}M_i^2+\gamma_{3\alpha}M_iL_i)P_\alpha^2,
\end{align}
where the components for $\vP$ run over $\alpha=(a,b,c)$ in the magnetic unit cell. The $\gamma_{1\alpha}$ and $\gamma_{2\alpha}$ terms have been introduced using symmetry arguments alone but their microscopic origin can e.g. originate from
\begin{eqnarray}
H_{int}=-\delta_{me}\sum_{ij}\sum_{kl}\vS_i\vS_j\sigma_k\sigma_l.
\end{eqnarray}
This Hamiltonian describes a bi-quadratic coupling between spins $\vS_{i,j}$ and structural pseudospins $\vsig_{k,l}$ \cite{Jiang:2002}. A bi-quadratic coupling term has also been derived by Pirc {\it et al.} \cite{Pirc:2010}. In multi-glass material like doped SrTiO$_3$ \cite{Shvartsmann:2008} or in EuTiO$_3$ \cite{Shvartsmann:2010} these terms have been invoked to explain the observed magnetoelectric effect. Finally, the DM induced bi-quadratic magnetoelectric coupling term with coefficient $\gamma_{3\alpha}$ has been used to explain magnetoelectricity in BaMnF$_4$ \cite{Fox:1980}.

The polarization free energy is given by
\begin{align}
F_P =  \sum_{\alpha}({\chi_{e\alpha}^{-1}\over 2}P_{\alpha}^2+{\beta\over 4}P_{\alpha}^4) -\vE\vP.
\end{align}
Here, $\chi_{e\alpha}$ is the electric susceptibility for the $\alpha$ component of the polarization. Fourth order terms have been included to obtain stable solutions, and $\vE$ denotes the applied electric field. In its most general form, the free energy should also contain gradient terms since we are dealing with a relaxor system \cite{Pirc:2010} as well as higher order terms. However, since we are interested in the magnetoelectric effect close to the ferroelectric transition, it is reasonable to restrict our analysis to the above terms.

The solutions that determine $\vP(\vH)$ are obtained by minimizing $F$ with respect to the electric polarization and magnetic order parameters. In the case of LCO studied experimentally, the N\'{e}el temperature is $T_N\sim 320$K, which is much higher than the temperature at which the ferroelectric order sets in ($T_P\sim 4.5$K) \cite{Zach:2011}. Therefore, we evaluate F$_M$ for the high temperature phase with $\vP=0$, providing the following set of equations
\begin{align}\label{Meq}
& M_i=\chi_0(H_c+CL_i),\\
& \chi_{2D}^{-1}L_1+AL_1^3+BL_1^5+{1\over 2}J_{\bot}L_2 = CM_1+H_{ab},\\
& \chi_{2D}^{-1}L_2+AL_2^3++BL_2^5+{1\over 2}J_{\bot}L_1 = CM_2+H_{ab},\\
& [\chi_{e\alpha}^{-1} +\sum_{i=1}^2(\gamma_{1\alpha}L_i^2+\gamma_{2\alpha}M_i^2+\gamma_{3\alpha}M_iL_i)]P_\alpha = -\beta P_\alpha^3.
\end{align}

\begin{figure}[b]
\begin{minipage}{.49\columnwidth}
\includegraphics[clip=true,width=0.98\columnwidth]{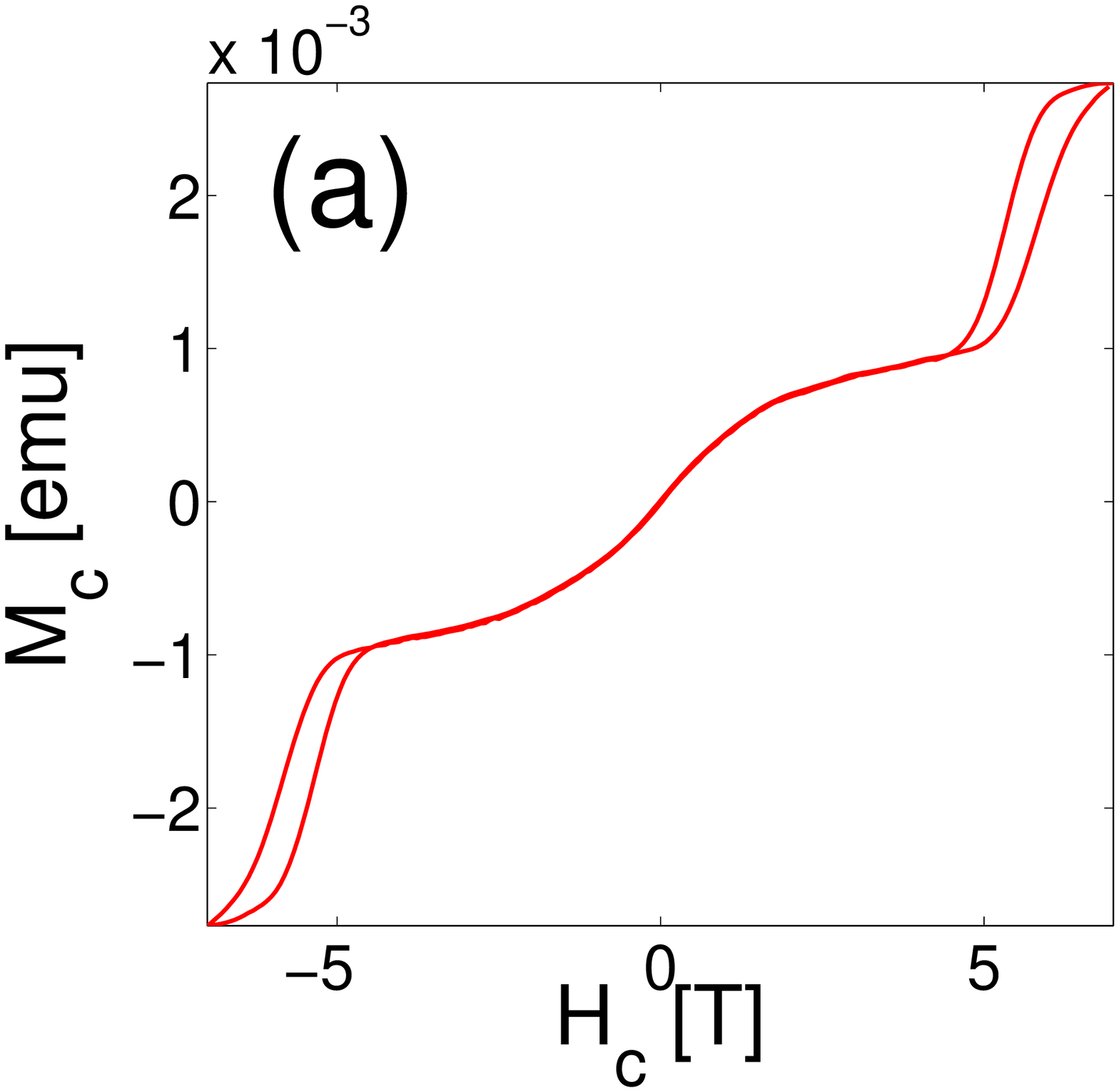}
\end{minipage}
\begin{minipage}{.49\columnwidth}
\includegraphics[clip=true,width=0.98\columnwidth]{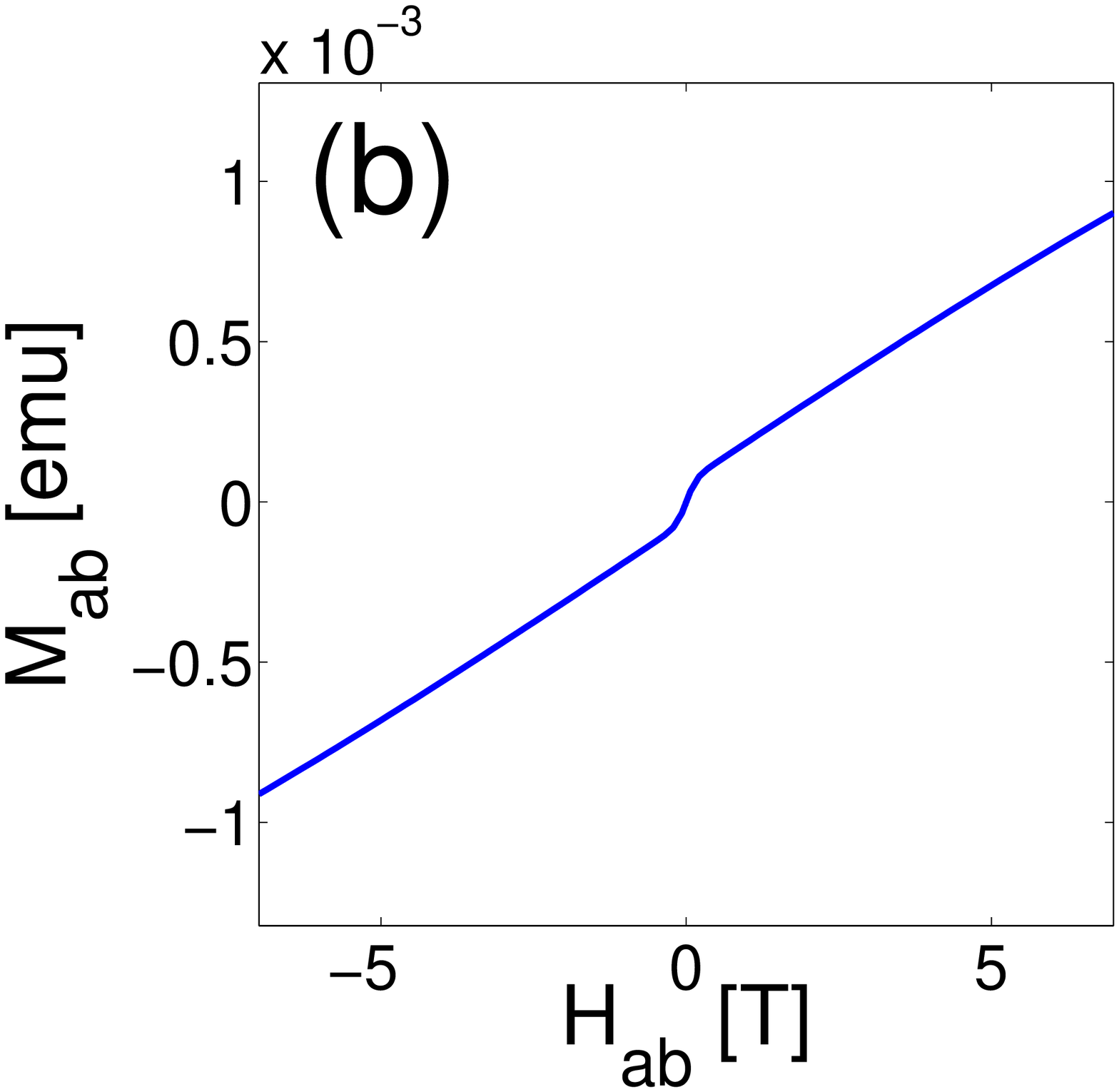}
\end{minipage}
\caption{(Color online) (a) Out-of-plane measured magnetization, and (b) in-plane magnetization measured at $T=5\rm{K}$. } \label{fig:magn}
\end{figure}

The experimental magnetization curves at low temperatures ($<\sim30$K) have a glassy contribution that can be observed in Fig.~\ref{fig:magn}(a) as a  hump feature at low magnetic fields and a broadened spin flop transition for out-of-plane magnetic fields $H_c$ \cite{Suzuki:2002}. Such glassy behavior is  absent in the magnetization response for in-plane applied magnetic fields $H_{ab}$ as seen from Fig.~\ref{fig:magn}(b). These features cannot be obtained from the above equations and hence to include them to the lowest order, we simply take the experimental magnetization curves in Fig.~\ref{fig:magn} as input. This approximation along with the temperature range at which we evaluate the magnetoelectric effect leads to deviations from the parameters of the magnetic free energy from those used e.g. by Thio {\it et al.}\cite{Thio:1994,coeff}

In terms of the following rescaled quantities $l_{+}=\chi_0C(L_1+L_2)/2$,
$l_{-}=\chi_0C(L_1-L_2)/2$, $M=(M_1+M_2)/2$, $\gamma_{1\alpha}'=2\gamma_{1\alpha}(\chi_0C)^{-2}$,
$\gamma_{2\alpha}'=2\gamma_{2\alpha}$, $\gamma_{3\alpha}'=2\gamma_{3\alpha}(\chi_0C)^{-1}$,
the polarization dependence on the applied magnetic field can be expressed as
\begin{align}\label{Pabeqn}
& {P_{\alpha}(H_{ab})\over P_{\alpha}(0)}=[1+{s_{\alpha}\over l_{-}^2(0)}(l_{-}^2(H_{ab})-l_{-}^2(0)+l_{+}^2(H_{ab}))]^{1/2},\\
& {P_{\alpha}(H_c)\over P_{\alpha}(0)}=[1+{s_{\alpha}\over l_{-}^2(0)}(l_{-}^2(H_c)-l_{-}^2(0)+g_{\alpha} M(H_c)^2\nonumber\\
& +(1-g_{\alpha}-q_{\alpha})l_{+}^2(H_c) +q_{\alpha} M(H_c)l_{+}(H_c))]^{1/2},
\end{align}
where $s_{\alpha}=\lambda_\alpha l_{-}^2(0)/ (\chi_{e\alpha}^{-1}+\lambda_\alpha l_{-}^2(0))$,
$g_\alpha={\gamma_{2\alpha}'/\lambda_{\alpha}}$, and $q_\alpha=\gamma_{3\alpha}'/\lambda_\alpha$ with
$\lambda_\alpha=\gamma_{1\alpha}'+\gamma_{2\alpha}'+\gamma_{3\alpha}'$. In general, all three parameters $s_\alpha$, $g_\alpha$, and $q_{\alpha}$ will be temperature dependent. The temperature dependence of $s_{\alpha}$ primarily results from its relation to the electric susceptibility; hence an estimation of the amplitude of $s_{\alpha}$ can reveal the magnitude of anisotropy in the electric polarization.

\begin{figure}[t]
\begin{minipage}{.49\columnwidth}
\includegraphics[clip=true,width=0.98\columnwidth]{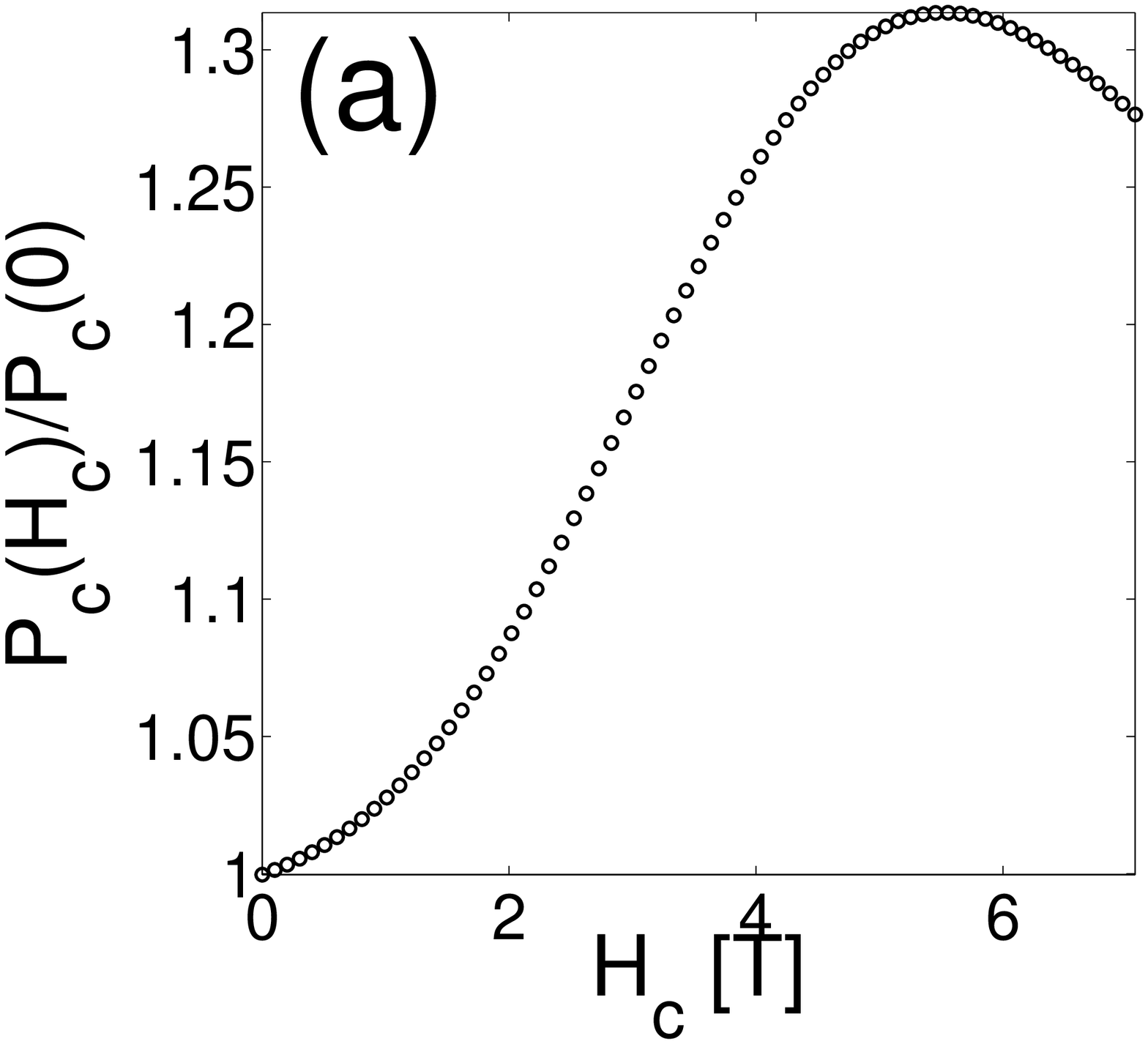}
\end{minipage}
\begin{minipage}{.49\columnwidth}
\includegraphics[clip=true,width=0.98\columnwidth]{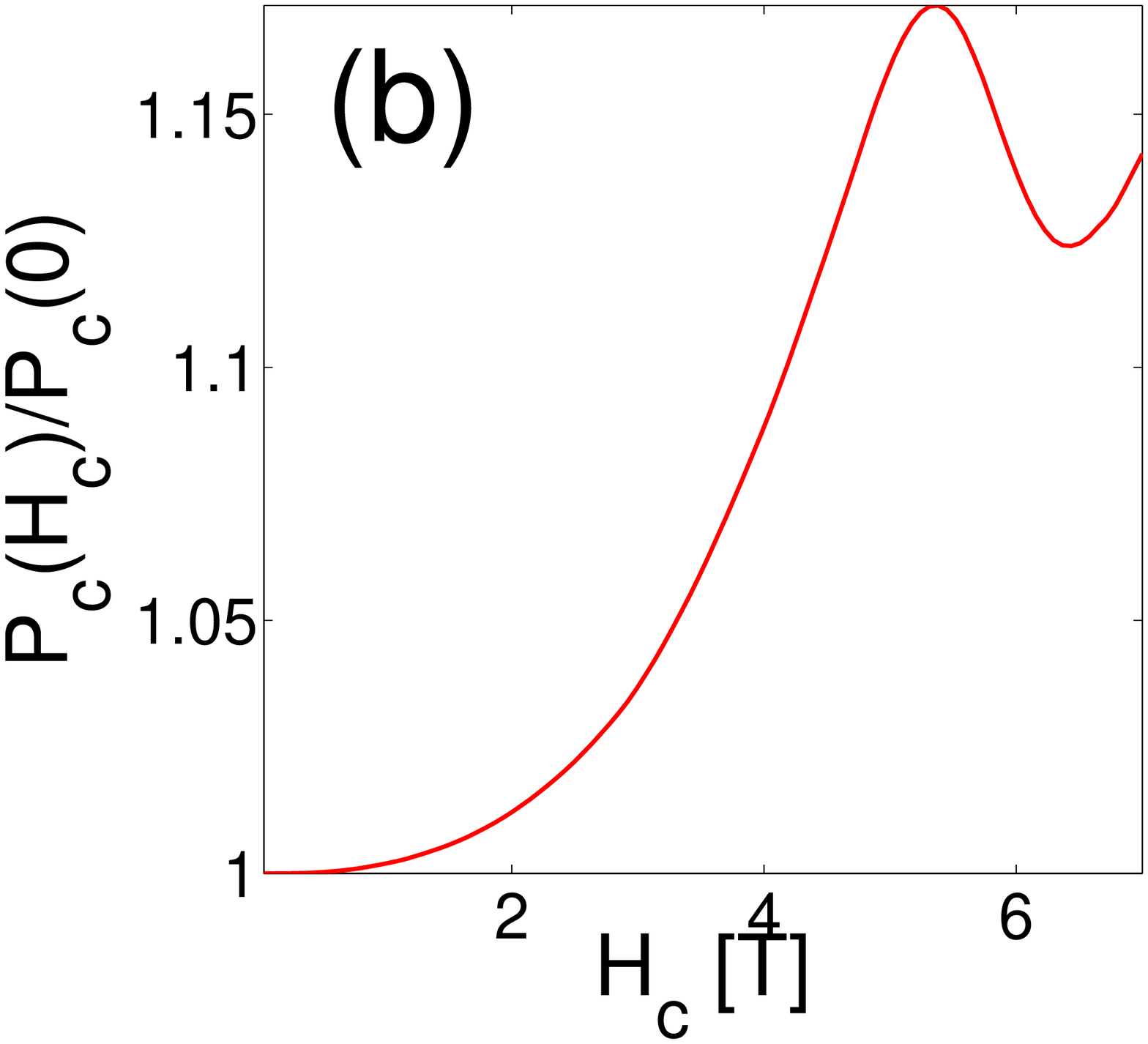}
\end{minipage}
\begin{minipage}{.49\columnwidth}
\includegraphics[clip=true,width=0.98\columnwidth]{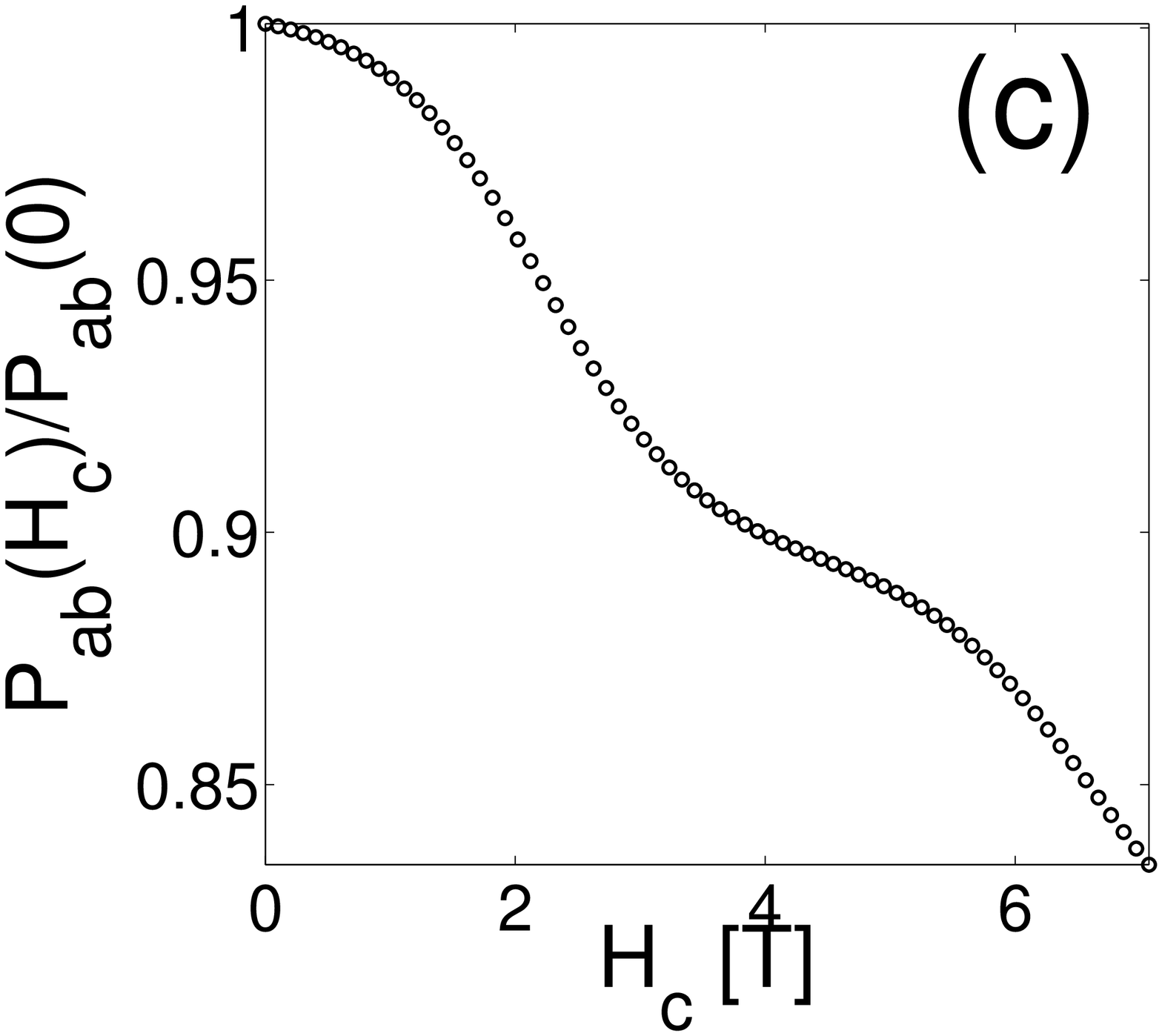}
\end{minipage}
\begin{minipage}{.49\columnwidth}
\includegraphics[clip=true,width=0.98\columnwidth]{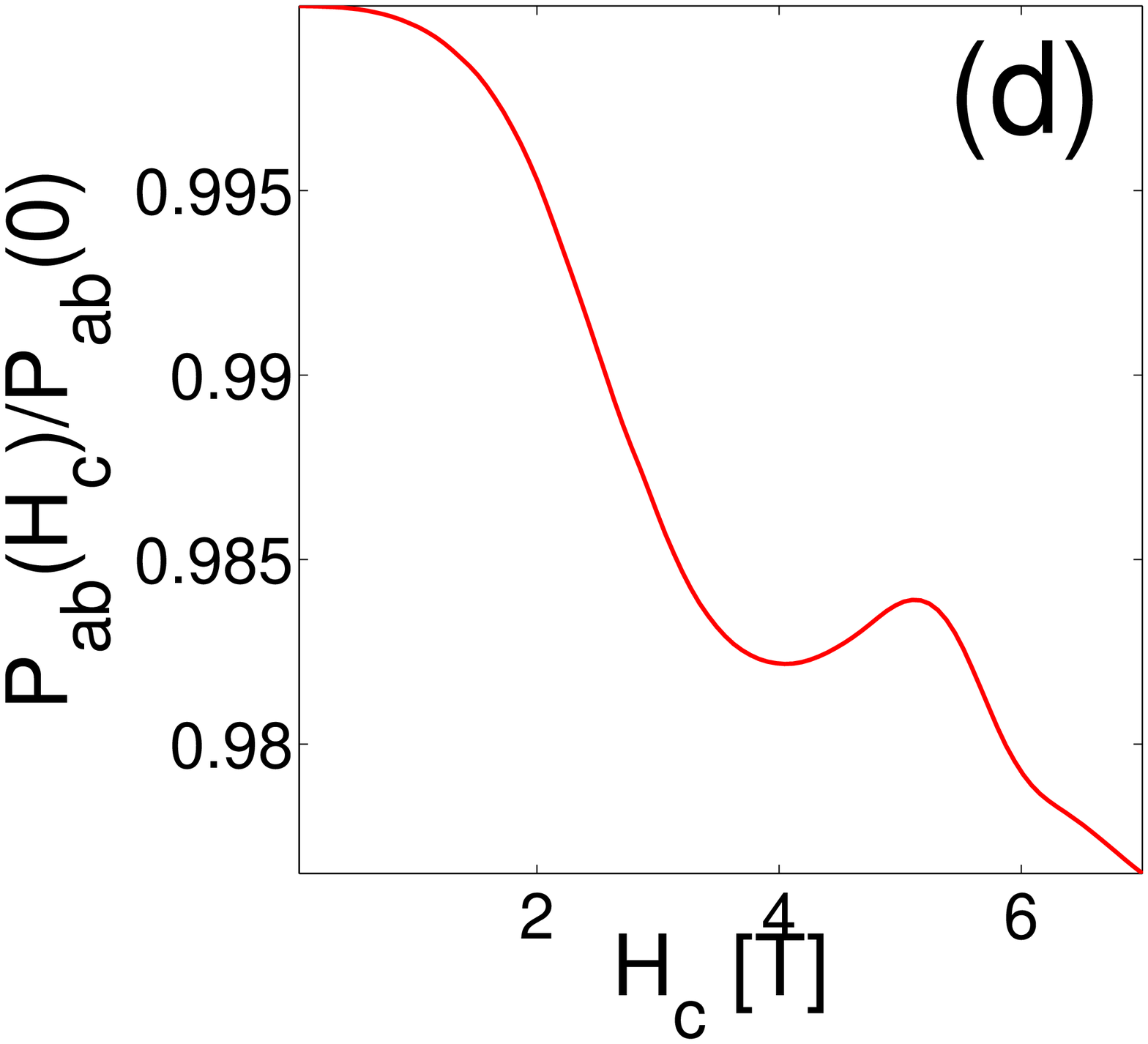}
\end{minipage}
\caption{(Color online) Polarization $P_{\alpha}$ vs. out-of-plane magnetic field $H_{c}$. (a,c) Experimentally measured values at $T=2.5\rm{K}$. (b,d) Theoretically calculated $P_{\alpha}(H_{c})/P_{\alpha}(0)$ for $s_c=0.1$, $s_a=0.074$, $g_a=g_b=g_c=-0.2$, $q_{c}=-6.58$, and $q_a=q_b=0$ using available experimental magnetization values at $T=5\rm{K}$. } \label{fig:pa}
\end{figure}

In the case of an out-of-plane magnetic field $H_c$, the measured $P_c(H_{c})$ increases with field and exhibits a pronounced hump at the spin-flop transition at $H_{\rm{sf}}$ as seen from Fig. \ref{fig:pa}(a). By contrast, $P_{ab}(H_{c})$ decreases with increasing $H_c$ but also exhibits a hump feature at $H_{\rm{sf}}$ as seen from Fig.~\ref{fig:pa}(c). This behavior indicates a scenario where the coupling between the magnetic order and out-of-plane polarization $P_c$ is attractive whereas the coupling with $P_{ab}$ is repulsive. However, we find that the theoretical picture is more complex due to the presence of two competing magnetic orders $\vL$ and $\vM$ coupling to the electric polarization. For the out-of-plane magnetic field, we calculate the $P_{\alpha}(H_{c})$ response using the magnetization data shown in Fig.~\ref{fig:magn}(a). As seen from Fig.~\ref{fig:pa}(b,d), we obtain qualitative agreement with both experimental polarization curves including the hump feature at $H_{\rm{sf}}$. Note that for the coupling to the magnetization we have set $g_a=g_b=g_c=-0.2$. Therefore, the source of anisotropy between $P_c(H_c)$ and $P_{ab}(H_c)$ is the DM induced coupling term $q_{\alpha}P_{\alpha}^2M_iL_i$ with $q_a=q_b=0$ and $q_c=-6.58$.

\begin{figure}[b]
\begin{minipage}{.49\columnwidth}
\includegraphics[clip=true,width=0.98\columnwidth]{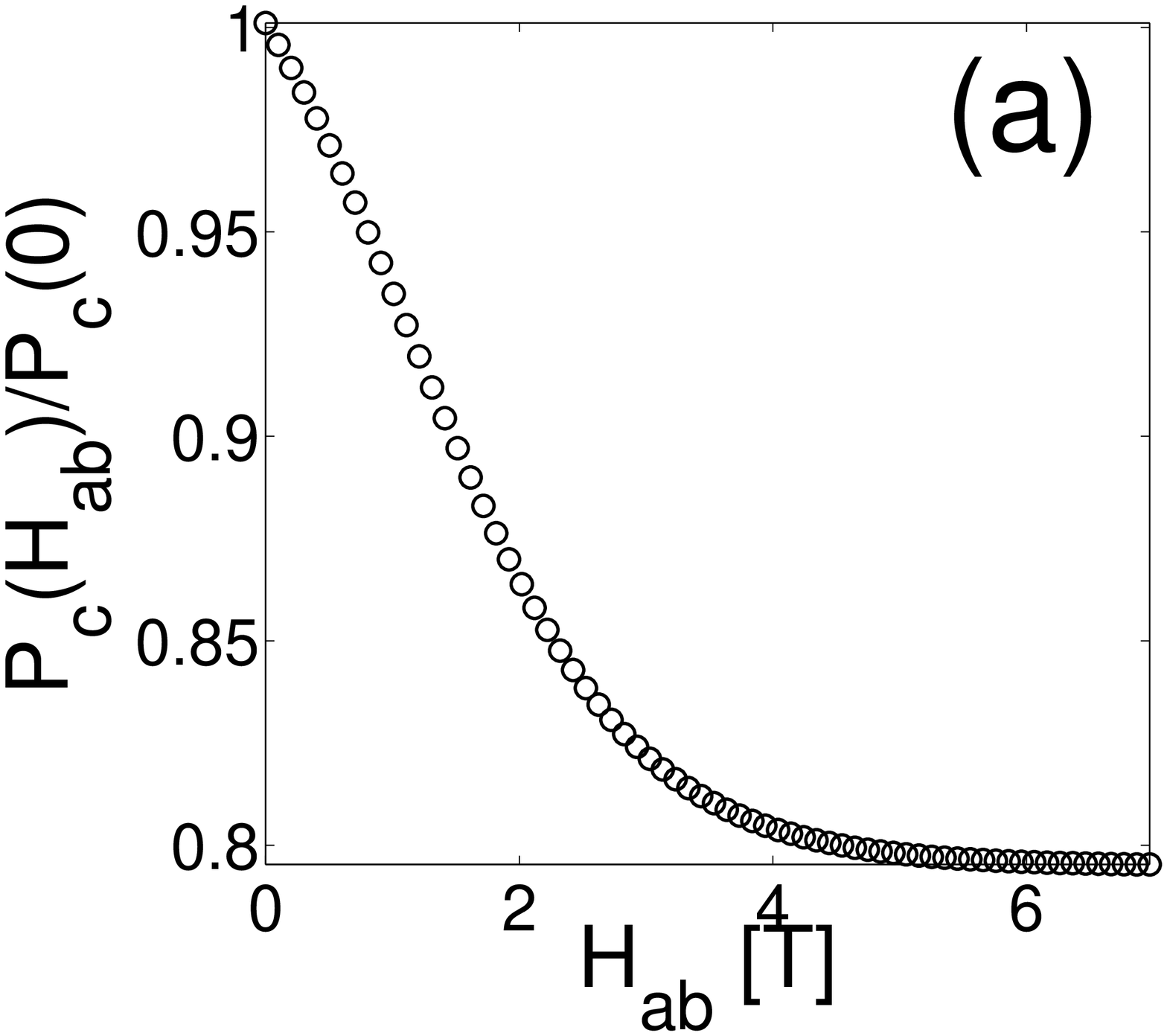}
\end{minipage}
\begin{minipage}{.49\columnwidth}
\includegraphics[clip=true,width=0.98\columnwidth]{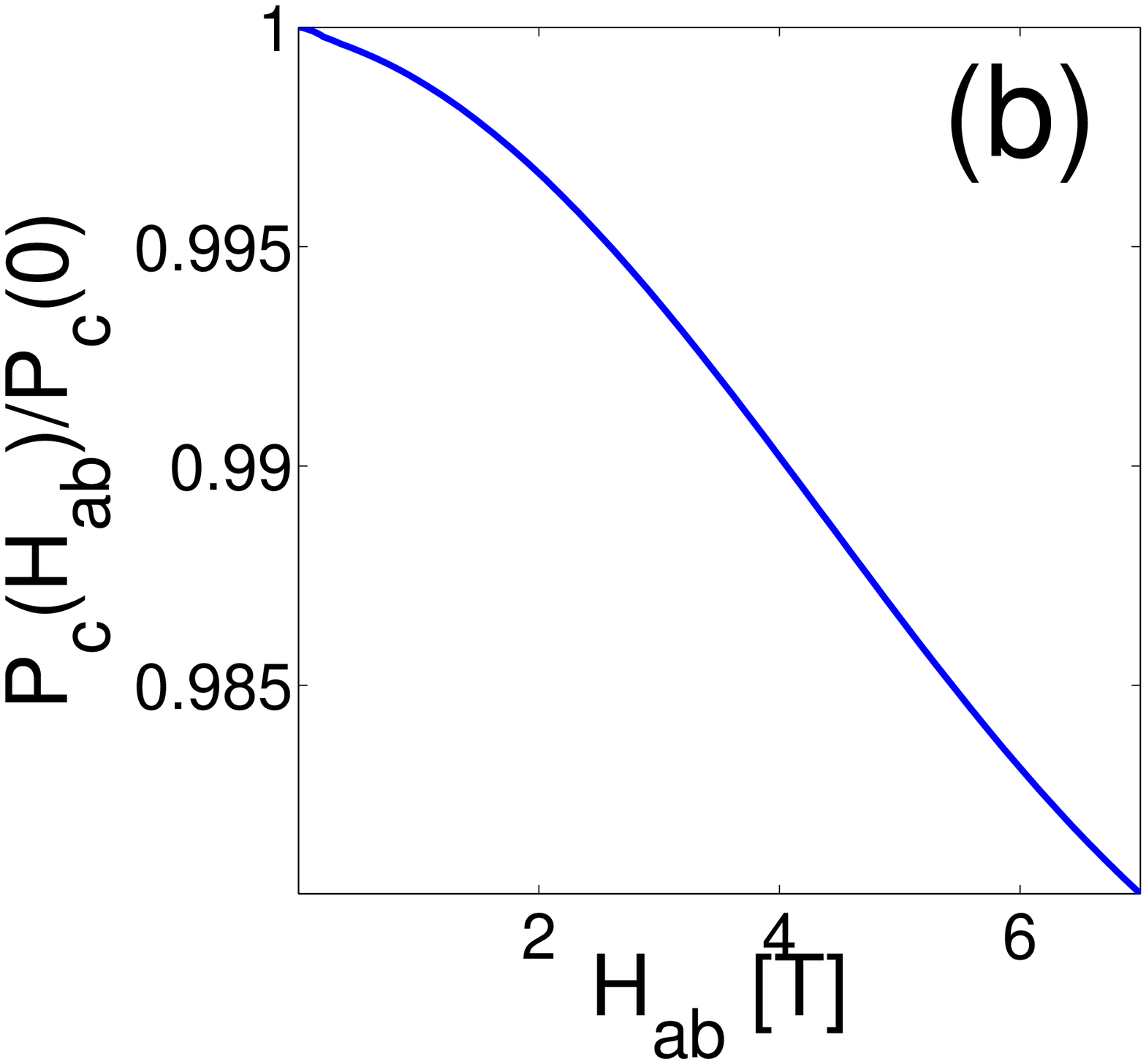}
\end{minipage}
\begin{minipage}{.49\columnwidth}
\includegraphics[clip=true,width=0.98\columnwidth]{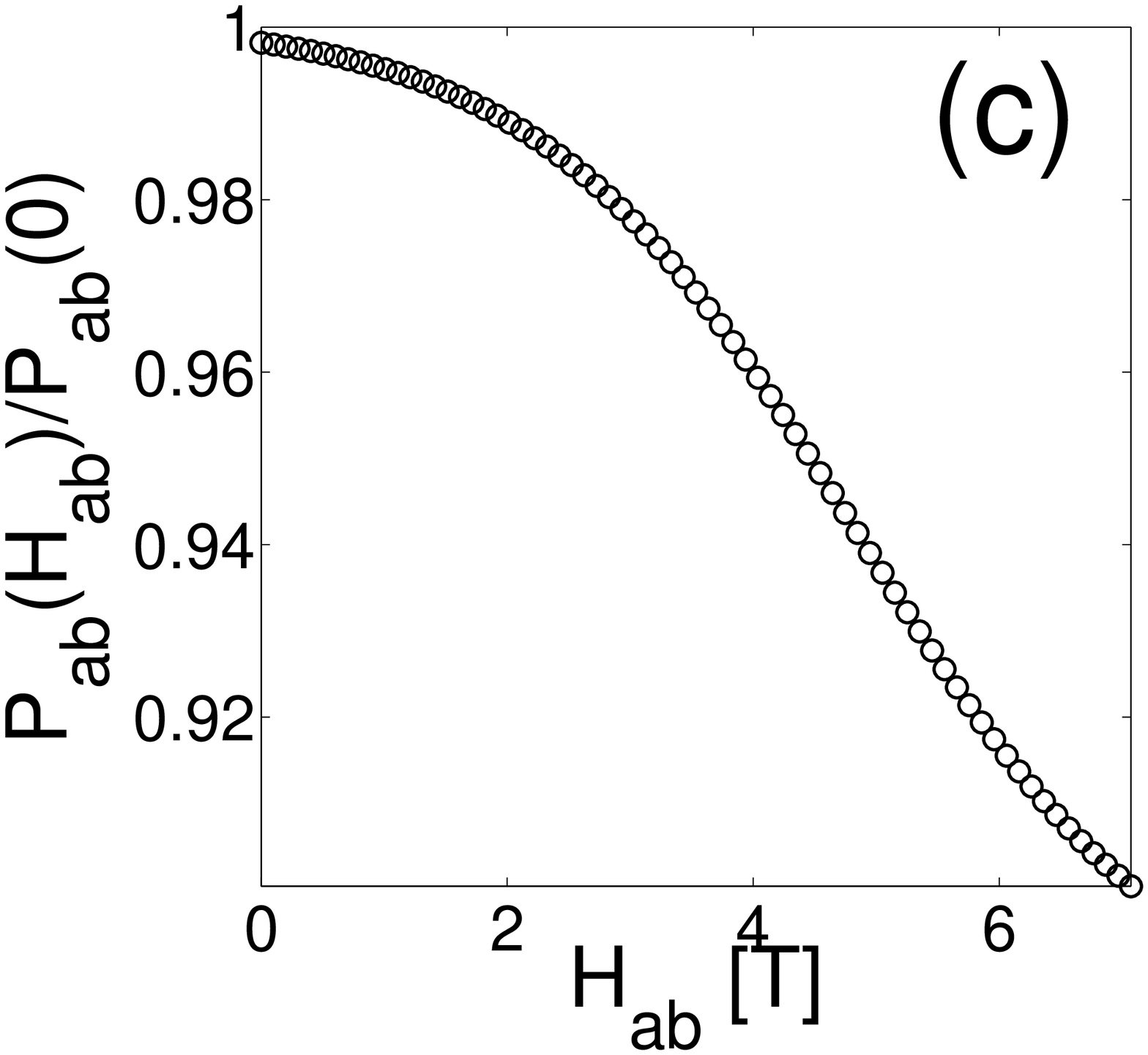}
\end{minipage}
\begin{minipage}{.49\columnwidth}
\includegraphics[clip=true,width=0.98\columnwidth]{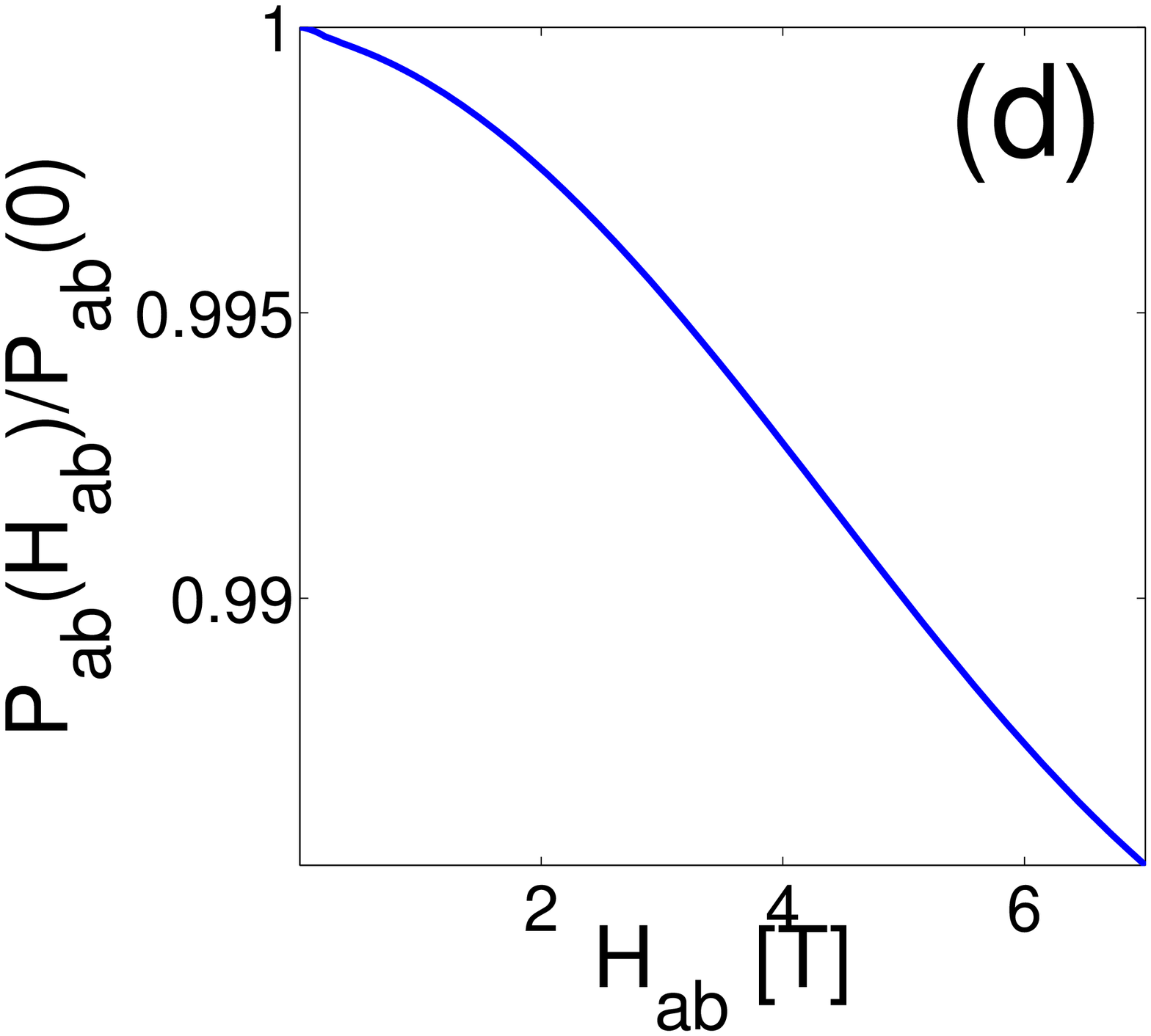}
\end{minipage}
\caption{(Color online) Polarization $P_{\alpha}$ vs. in-plane magnetic field $H_{ab}$. (a,c) Experimentally measured values at $T=2.5\rm{K}$. (b,d) Theoretically calculated  $P_{\alpha}(H_{ab})/P_{\alpha}(0)$ for $s_c=0.1$ and $s_a=0.074$ using available experimental magnetization values at $T=5\rm{K}$.} \label{fig:pc}
\end{figure}

In the case of an in-plane magnetic field $H_{ab}$, $P_{\alpha}(H_{ab})$ depends on only a single fitting parameter $s_{\alpha}$ that controls the magnitude of the polarization ratio, whereas the shape of the theoretical curves are governed by the magnetic order parameter of the system. Further, we can deduce from Eq.~\eqref{Pabeqn} that the polarization primarily couples to the 2D antiferromagnetic order through a magnetoelectric interaction term $F_{MP}\sim\lambda_{\alpha}P_{\alpha}^2L_i^2$. The anisotropy in polarization through the parameter $\lambda_{\alpha}$ for in-plane magnetic fields is primarily controlled by the DM induced magnetoelectric coupling term $q_{\alpha}P_{\alpha}^2M_iL_i$, since $q_{\alpha}$ undergoes the largest variation between in-plane and out-of-plane directions. The DM physics therefore plays an important role in generating an anisotropy between $P_c(\vH)$ and $P_{ab}(\vH)$ in La$_2$CuO$_{4+x}$.

Using the magnetization curve from Fig. \ref{fig:magn}(b), the solutions of $P_{\alpha}(H_{ab})$ are presented in Fig. \ref{fig:pc}(b,d). Though the theoretical curves have similar shapes, the scales are different in the two figures due to the difference in $s_{\alpha}$ values. Note that all theoretical curves have been plotted using magnetization data available at T=5K while the experimental curves correspond to T=2.5K which accounts for the smaller scale of the theoretical curves. As seen from the plots, we find again qualitative agreement with the experimental data shown in Figs. \ref{fig:pc}(a,c) though the experimental plot in \ref{fig:pc}(a) has a steeper slope than theory. As in Fig.~\ref{fig:pa}, the plots correspond to $s_c=0.1$, $s_a=0.074$ which leads to a small anisotropy in the electric susceptibilities $\chi_{e\alpha}^{-1}$ and therefore between the zero magnetic field values of in-plane polarization $P_{ab}$ and the out-of-plane component $P_c$. This small anisotropy has been observed in experiments \cite{Zach:2011} and implies a much weaker anisotropy of the electric polarization compared to the magnetic order. This result lends support to a scenario where the non-stoichiometric oxygen dopants (charge carrier doping) play an important role in generating the relaxor ferroelectricity in La$_2$CuO$_{4+x}$.

We have observed experimentally that the magnetization shows a small enhancement below the temperatures where the ferroelectric order sets in \cite{Zach:future}. This effect is in addition to the typical upturn in magnetization near the spin glass freezing temperature \cite{Keimer:1992,Matsuda:2002}. We can study such feedback effect of a finite polarization on the magnetization by minimizing $F$ within the ferroelectric phase. This gives to lowest order
\begin{align}
M_{c}={\chi_0H_c+[1-\chi_0\sum_{\alpha}\gamma_{3\alpha}'P_{\alpha}^2(H_c)]l_{+}(H_c)\over 1+\chi_0\sum_{\alpha} \gamma_{2\alpha}'P_{\alpha}^2(H_c)},\\
M_{ab}={[1-\chi_0\sum_{\alpha}\gamma_{3\alpha}'P_{\alpha}^2(H_{ab})]l_{+}(H_{ab})\over 1+\chi_0\sum_{\alpha} \gamma_{2\alpha}'P_{\alpha}^2(H_{ab})}.
\end{align}
Note that in this expression the relative sign of the coefficients can be determined from the relations $\gamma_{3\alpha}'/\gamma_{2\alpha}'=q_\alpha/g_\alpha>0$. This also implies that since $\gamma_{2\alpha}'<0$ in our model, it naturally causes an enhancement of magnetization due to the presence of a ferroelectric state. Additionally, we also find that this enhancement is present both in the in-plane and out-of-plane magnetization with the relative size of the enhancement depending on the amplitude of polarization change with magnetic field.

A bi-quadratic coupling has been used to explain the observation of magnetocapacitive effects in materials like doped SrTiO$_3$  \cite{Shvartsmann:2010}. It is defined by the relation $\epsilon_\alpha=-\partial^2 F/ \partial P_\alpha^2$ and hence requires at least quadratic terms in the polarization. For LCO the relative change in dielectric constant is given by
\begin{eqnarray}
\Delta \epsilon_{\alpha} = [\epsilon_{\alpha}(\vH)-\epsilon_{\alpha}(0)]/ \epsilon_{\alpha}(0)=\Delta P_\alpha^2(\vH).
\end{eqnarray}

A weak magnetocapacitive effect is therefore predicted in experiments at low temperatures. Note that in the above expression we would expect a small suppression in the permittivity for magnetic fields in the $a-b$ plane.

\begin{figure}
\includegraphics[clip=true,width=0.6\columnwidth]{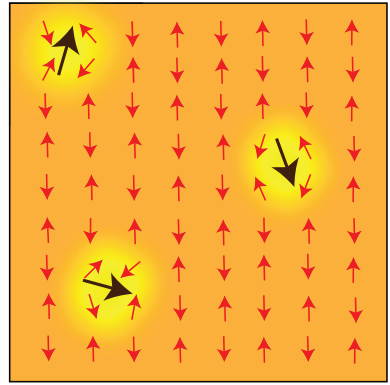}
\caption{(Color online) Illustration of polar nano-regions with randomly oriented electric polarizations (black arrows) at high $T$ within the antiferromagnet (red arrows). Also shown are the distorted magnetic moments within a correlation length of the polar nano-regions.}
\label{fig:pnr}
\end{figure}

The observation of relaxor ferroelectricity in underdoped LCO has been argued to originate from the formation of polar nano-regions (PNR) around the non-stoichiometric oxygen dopants\cite{Zach:2011}. The formation of PNR and the mechanism by which they condense into a ferroelectric phase is a well studied topic \cite{Samara:2003,Guangyong:2006,Guangyong:2008}. Though the relaxor physics in LCO naturally relates to the presence of dopants, the extremely low concentration of excess oxygen in the samples used in Ref.~\onlinecite{Zach:2011} may imply the presence of additional mechanisms for the PNR to couple and undergo a spontaneous transition to long-range ferroelectric order. One may speculate that such mechanisms include subtle non-centrosymmetric structural distortions in the host lattice\cite{Reehuis:2006} and/or a tendency for the dopants to cluster and thereby reduce the inter-PNR distance.  As shown in Fig.~\ref{fig:pnr}, in magnetic materials like LCO the PNRs may also cause a distorted spin structure that could lead to a magnetoelectric effect through e.g. geometric frustrations in the presence of DM interaction and/or indirectly through coupling to strains. This physics has similarities to the observation of magnetoelectric behavior in a number of other relaxor ferroelectrics\cite{Shvartsmann:2008, Nugroho:2007}.

In summary, we have shown that the magnetoelectric effect in extremely underdoped La$_2$CuO$_{4+x}$ can be explained by bi-quadratic terms in the free energy. It is proposed that the microscopic origin of the ferroelectricity is caused by polar nano-regions generated by dopant ions. The discovery of ferroelectricity and magnetoelectric effect in the cuprate materials due to charge carrier doping have sparked many questions for future studies. In particular, what happens at higher doping levels and what is the fate of the PNRs in the regime where the pseudo-gap and superconducting phases emerge?

We acknowledge the financial support by the European Union through MEXT-CT-2006-039047 and EURYI research grants.  The work in Singapore was funded by The National Research Foundation.  B.M.A. acknowledges support from The Danish Council for Independent Research $|$ Natural Sciences.


\begin{thebibliography}{100}

\bibitem{Eerenstein:2006} W. Eerenstein, N. D. Mathur, and J. F. Scott, Nature (London) {\bf 442}, 759 (2006).
%
\bibitem{Cheong:2007} S.-W. Cheong and M. Mostovoy, Nature Mater. {\bf6}, 13 (2007).
%
\bibitem{Levstik:2007} A. Levstik {\it et al.}, Appl. Phys. Lett. {\bf 91}, 012905, (2007).
%
\bibitem{Zach:2011} Z. Viskadourakis {\it et al.}, arXiv:1111.0050v1.
%
\bibitem{Cross:1987} L. E. Cross, Ferroelectrics {\bf 76}, 241 (1987).
%
\bibitem{Samara:2003} G. A. Samara, J. Phys. Condens. Mater. {\bf 15}, R367 (2003).
%
\bibitem{Kastner:1998} M. A. Kastner, R. J. Birgeneau, G. Shirane, and Y. Endoh, Rev. Mod. Phys. {\bf 70}, 897 (1998).
%
\bibitem{Keimer1:1992} B. Keimer {\it et al.}, Phys. Rev. B {\bf 46}, 14034 (1992).
%
\bibitem{Thio:1988} T. Thio {\it et al.},  Phys. Rev. B {\bf 38}, 905 (1988).
%
\bibitem{Reehuis:2006} M. Reehuis {\it et al.},  Phys. Rev. B {\bf 73}, 144513 (2006).
%
\bibitem{Thio:1990} T. Thio {\it et al.},  Phys. Rev. B {\bf 41}, 231 (1990).
%
\bibitem{Lavrov:2002} A. N. Lavrov, S. Komiya, and Y. Ando, Nature (London) {\bf 418}, 385 (2002).
%
\bibitem{Thio:1994} T. Thio and A. Aharony, Phys. Rev. Lett. {\bf 73}, 894 (1994).
%
\bibitem{Dzyaloshinskii:1991} I. E. Dzyaloshinskii, Phys. Lett. A {\bf 155}, 62 (1991).
%
\bibitem{Jiang:2002} Q. Jiang and H. Wu, Chin. Phys. {\bf 11}, 1303 (2002).
%
\bibitem{Pirc:2010} R. Pirc and R. Blinc, Ferroelectrics {\bf 400}, 387 (2010).
%
\bibitem{Shvartsmann:2008} V. V. Shvartsmann {\it et al.}, Phys. Rev. Lett. {\bf 101}, 165704 (2008).
%
\bibitem{Shvartsmann:2010} V. V. Shvartsmann {\it et al.}, Phys. Rev. B. {\bf 81}, 064426 (2010).
%
\bibitem{Fox:1980} D. L. Fox, D. R. Tilley, and J. F. Scott, Phys. Rev. B. {\bf 21}, 2926 (1980).
%
\bibitem{Suzuki:2002} T. Suzuki {\it et al.}, Phys. Rev. B. {\bf 66}, 172410 (2002).
%
\bibitem{coeff} The coefficients in $F_M$ are $\chi_0=5.625\times 10^{-4}\rm{cm}^3/\rm{mole}$, $C\chi_0\!=\!1.88\times 10^{-3}$, $\chi_{2D}^{-1}-C^2\chi_0-J_{\bot}=-4.4\mu \rm{eV}$, $a=2.4\times 10^{-3}/(C\chi_0)^2\rm{eV}/(\rm{emu})^2$, $b=10^{-3}/(C\chi_0)^4\rm{eV}/(\rm{emu})^2$.
%
\bibitem{Zach:future} Z. Viskadourakis, {\it private communication}.
%
\bibitem{Keimer:1992} B. Keimer {\it et al.},  Phys. Rev. B {\bf 45}, 7430 (1992).
%
\bibitem{Matsuda:2002} M. Matsuda {\it et al.},  Phys. Rev. B {\bf 65}, 134515 (2002).
%
\bibitem{Guangyong:2006}X. Guangyong {\it et al.}, Nature Mater. {\bf 5}, 134 (2006).
%
\bibitem{Guangyong:2008}X. Guangyong {\it et al.}, Nature Mater. {\bf 7}, 562 (2006).
%
\bibitem{Nugroho:2007} A. A. Nugroho {\it et al.},  Phys. Rev. B {\bf 75}, 174435 (2007).
%
\end{thebibliography}
\end{document}